\documentstyle[12pt]{article}

\newlength{\HFPP}       \HFPP5.4mm
\makeatletter
\@addtoreset{equation}{section}
\makeatother

\newcommand{\beqn}{\begin{equation}}
\newcommand{\eeqn}{\end{equation}}
\newcommand{\beqa}{\begin{eqnarray}}
\newcommand{\eeqa}{\end{eqnarray}}
\newcommand{\nn}{\nonumber\\}

\begin{document}

\begin{titlepage}

\begin{flushleft}
\setlength{\baselineskip}{13pt}
Yukawa Institute Kyoto \hfill YITP-96-33 \\
		       \hfill August 1996
\end{flushleft}

\vspace*{\fill}
\begin{center}
{\Large Scattering of Plane Waves 
                in Self-Dual Yang-Mills Theory} \\
\vfill
{\sc Vladimir E. Korepin$^{1-3}$
\footnote{e-mail: {\tt korepin@insti.physics.sunysb.edu}}
and 
Takeshi Oota$^3$}
\footnote{e-mail: {\tt toota@yukawa.kyoto-u.ac.jp}}\\[2em]

$^1${\sl Institute for Theoretical Physics, State University of New
           York at Stony Brook,\\
           Stony Brook, NY 11794--3840, U. S. A.}\\
$^2${\sl Sankt Petersburg Department of Mathematical Institute of
           Academy of Sciences of Russia}
$^3${\sl Yukawa Institute for Theoretical Physics, Kyoto
           University, Kyoto 606-01, Japan}\\

\vfill
ABSTRACT
\end{center}
\begin{quote}
We consider the classical self-dual Yang-Mills equation 
in 3+1-dimensional Minkowski space. We have found an exact solution, 
which describes scattering of $n$ plane waves. 
In order to write the solution in a compact form, 
it is convenient to introduce a scattering operator $\hat{T}$. 
It acts in the direct product of three linear spaces: 
1) universal enveloping of $su(N)$ Lie algebra, 
2) $n$-dimensional vector space and 
3) space of functions defined on the unit interval.
\end{quote}
\vfill
\setcounter{footnote}{0}
\end{titlepage}


\section{Introduction}
We consider the classical Yang-Mills field with value in $su(N)$ algebra
in 3+1-dimensional Minkowski space. We study the self-dual equation:
\beqn
  F_{\mu \nu} 
  = \frac{i}{2} \epsilon_{\mu \nu \rho \sigma} F^{\rho \sigma}.
\eeqn
Study of the self-dual Yang-Mills equation is important 
for understanding of QCD \cite{P}\nocite{BPST,ADHM,KS,T,dV}--\cite{W}.

We fix the gauge following \cite{BLR}:
\beqn
  A_{0 - z} = 0, \ \ \ A_{x+iy} = 0, \ \ \ 
  A_{0 + z} = \sqrt{2} \partial_{x + iy} \Phi, \ \ \ 
  A_{x - iy} = \sqrt{2} \partial_{0 - z} \Phi.
\eeqn
Here $A_{0 \pm z} = A_0 \pm A_z$, $A_{x \pm iy} = A_x \mp i A_y$.

The self-dual Yang-Mills equation for scalar $su(N)$ field $\Phi$ looks as
follows
\beqn
  \Box \Phi - ig[ \partial_{x + iy} \Phi, \partial_{0 - z} \Phi ] = 0.
\label{SD}
\eeqn
It is associated with a cubic action \cite{LMP}. Following \cite{C}
we start looking for the solution of eq.(\ref{SD}) using perturbation
theory in coupling constant $g$
\beqn
  \Phi(x) = \sum_{m = 1}^{\infty} \Phi^{(m)}(x).
\label{exg}
\eeqn
Here $\Phi^{(m)}$ depends on the coupling as $g^{m-1}$. 
The first term satisfies a linear equation
\beqn
  \Box \Phi^{(1)} = 0.
\eeqn
We choose $\Phi^{(1)}$ as a sum of $n$ plane waves
\beqn
  \Phi^{(1)}(x) = -i\sum_{j=1}^n T^{a_j} e^{-ik_j x} f(k_j).
\eeqn
Here $T^a$ are $su(N)$ generators
\beqa
  [ T^a, T^b ] &=& i\sqrt{2} f^{abc} T^c, \nn
  tr T^a T^b &=& \delta^{ab}.
\eeqa
All $k_j$ are $n$ different light-cone vectors $k_j^2=0$ and $f(k)$ is 
a function with support on the light-cone. We will also use the
following notations:
\beqn
  Q_j = \frac{(k_j)_{0 + z}}{(k_j)_{x + iy}}
      = \frac{(k_j)_{x - iy}}{(k_j)_{0 - z}}.
\eeqn
We have found explicit expression for $\Phi^{(m)}$. The first two terms
coincide with \cite{C}, all other $\Phi^{(m)}$ $(m \geq 3)$ are
different. 

Let us explain our solution. 

We shall use an abbreviation:
\beqn
  \phi(j) = T^{a_j} e^{-ik_j x}f(k_j).
\label{fp}
\eeqn

We introduce the following function:
\beqn
  V(a) = \sum_{n = 0}^{\infty} \frac{1}{(n!)^2}a^n
       = \oint \frac{dt}{2\pi i}\frac{e^{1/t + at}}{t}
       = I_0(2\sqrt{a}).
\label{mB0}
\eeqn
Here $I_{0}$ is the modified Bessel function of the first kind.
The integration contour is a circle around zero. We integrate in positive 
direction.

Let us define a linear operator:
\beqa
  & &T(\alpha_1, \alpha_2; j_1, j_2) \nn
  &=&
    g \phi(j_1) P(j_1, j_2)
       \int_0^{\infty}ds\ e^{-s}
         V\Bigl( s \alpha_1 g \phi(j_1) P(j_1, j_2) \Bigr)
         V\Bigl( s \alpha_2 g \phi(j_2) P(j_1, j_2) \Bigr). 
\label{scop}
\eeqa
Here $j_1$ and $j_2$ run through $n$ values. Integration variables 
$\alpha_{1, 2}$ belong to a unit interval $[0, 1]$.
We shall consider $T$ as an operator in direct product of
$n$-dimensional vector space and as an integral operator. The kernel 
$T(\alpha_1, \alpha_2; j_1, j_2)$ takes values in an universal
enveloping algebra of $su(N)$. We are using $P(j_1, j_2)$ which is
defined by
\beqn
 P(j_1, j_2) 
    = \left\{\begin{array}{ccc}
        ( Q_{j_1} - Q_{j_2} )^{-1} & & {\rm for}\ \ j_1 \neq j_2 \\
                                   & & \\
                                 0 & & {\rm for}\ \ j_1 = j_2.
             \end{array}\right.
\label{P12}
\eeqn

The kernel $T(\alpha, \alpha'; j, j')$ depends only 
on the $j$-th and $j'$-th plane waves. It vanishes if $j = j'$.

The function (\ref{scop}) is a kernel of an operator
 $\hat{T}$ 
\[
  (\hat{T})_{(\alpha_1; j_1), (\alpha_2; j_2)}
  = T(\alpha_1, \alpha_2; j_1, j_2),
\]
whose index $(\alpha; j)$ takes a value 
in $[0,1]\times\{1, 2, \ldots, n\}$. 
It acts on a ``vector'' $(\vec{f})_{(\alpha; j)}$ which takes a value
in the universal enveloping algebra as follows:
\beqn
  ( \hat{T} \vec{f} )_{(\alpha; j)}
  = \sum_{j' = 1}^n \int_0^1 d\alpha' T(\alpha, \alpha'; j, j') 
     (\vec{f})_{(\alpha'; j')}.
\eeqn
We call $\hat{T}$ the `scattering operator'.

We introduce two special ``vector''s (see (\ref{fp}))
\beqn
  (\vec{\phi}  )_{(\alpha; j)} = \phi(j), \ \ \ 
  (\vec{\phi}_0)_{(\alpha; j)} = 1.
\label{tvec}
\eeqn
For example, a scalar product of $\vec{\phi}_0$ and an arbitrary vector
function $\vec{f}$ is equal to:
\[
  \vec{\phi}_0 \cdot \vec{f} 
   = \sum_{j = 1}^n \int_0^1 d\alpha \ (f)_{(\alpha; j)}.
\]
Now all the notations are prepared in order to write down 
the solution of the self-dual equation (\ref{SD}) which we have found:
\beqn
  \Phi(x)
  = -i \vec{\phi}_0 \cdot \left( \frac{1}{1 - \hat{T}} \right) \vec{\phi}.
\label{main}
\eeqn
This is the main result of our paper.

\section{Solving recursion relations}

Substituting eq.(\ref{exg}) into the self-dual equation (\ref{SD}) 
gives the following recursion relations for $\Phi^{(m)}(x)$:
\beqn
\label{Phrec}
  \Box \Phi^{(m)} (x)
  = ig \sum_{j = 1}^{m-1}
    ( \partial_{x + iy} \Phi^{(j)} \partial_{0 - z} \Phi^{(m - j)}
       - \partial_{0 - z} \Phi^{(j)} \partial_{x + iy} \Phi^{(m - j)} ).
\eeqn
It is convenient to introduce the following object:
\beqn
  X(1, 2) = (k_1)_{x + iy} (k_2)_{0 - z}-(k_1)_{0 - z} (k_2)_{x + iy}.
\label{X12}
\eeqn
It has the property:
\beqn
  2k_1 \cdot k_2 = X(1, 2) ( Q_1 - Q_2 ).
\eeqn
The lowest component of $\Phi(x)$ in (\ref{exg}) satisfies the free equation:
\beqn
 \Box \Phi^{(1)}=0.
\eeqn
A solution is given by
\beqn
  \Phi^{(1)}(x) = -i\sum_{j = 1}^n 
                  T^{a_j} e^{-ik_j x} f(k_j), \ \ \ \ k_j^2=0.
\eeqn
We shall use the notation:
\beqn
  \phi(j) = T^{a_j} e^{-ik_j x} f(k_j).
\label{phj}
\eeqn
Let us analyze  $\Phi^{(2)}(x)$. The recursion relation (\ref{Phrec}) is
\beqa
  \Box \Phi^{(2)}(x)
  &=& ig ( \partial_{x+iy} \Phi^{(1)} \partial_{0-z} \Phi^{(1)}
      -\partial_{0-z} \Phi^{(1)} \partial_{x+iy} \Phi^{(1)}) \nn
  &=& ig \sum_{j_1}\sum_{j_2}
       \left( \partial_{x+iy} \phi(j_1) \partial_{0-z} \phi(j_2)
           -\partial_{0-z} \phi(j_1) \partial_{x+iy} \phi(j_2)\right) \nn
  &=& -ig \sum_{j_1}\sum_{j_2} \phi(j_1)\phi(j_2)X(j_1, j_2) \nn
  &=&  ig \sum_{j_1\neq j_2} \phi(j_1)\phi(j_2)X(j_1, j_2).
\label{Phtwo}
\eeqa
Here we have used the property $X(j_1, j_1)=0$.

Equation (\ref{Phtwo}) can be solved by:
\beqa
  \Phi^{(2)}(x)
  &=& -ig \sum_{j_1} \sum_{j_2(\neq j_1)}
      \phi(j_1) \phi(j_2) X(j_1, j_2)/(k_{j_1} + k_{j_2})^2 \nn
  &=& -ig \sum_{j_1}\sum_{j_2(\neq j_1)}
      \phi(j_1) \phi(j_2) (Q_{j_1} - Q_{j_2})^{-1}.
\label{min2}
\eeqa
In general, we can add any terms which satisfy the free equation 
to this solution $\Phi^{(2)}(x)$. 
We choose the minimal solution (\ref{min2}).

Next, we consider an equation for $\Phi^{(3)}(x)$.
\beqa
  \Box \Phi^{(3)}(x)
  &=& ig
      ( \partial_{x + iy} \Phi^{(1)} \partial_{0 - z}  \Phi^{(2)}
      - \partial_{0 - z}  \Phi^{(1)} \partial_{x + iy} \Phi^{(2)}) \nn
  &+& ig
      ( \partial_{x + iy} \Phi^{(2)} \partial_{0 - z}  \Phi^{(1)}
      - \partial_{0 - z}  \Phi^{(2)} \partial_{x + iy} \Phi^{(1)}) \nn
  &=& ig^2 \sum_{j_1} \sum_{j_2} \sum_{j_3(\neq j_2)}
      \phi(j_1) \phi(j_2) \phi(j_3) (Q_{j_2} - Q_{j_3})^{-1}
        \left( X(j_1, j_2) + X(j_1, j_3) \right) \nn
  &+& ig^2 \sum_{j_1} \sum_{j_2(\neq j_1)} \sum_{j_3}
      \phi(j_1) \phi(j_2) \phi(j_3) (Q_{j_1} - Q_{j_2})^{-1}
        \left( X(j_1, j_3) + X(j_2, j_3) \right) \nn
  &=& ig^2 \sum_{j_1} \sum_{j_2(\neq j_1)} \sum_{j_3(\neq j_2)}
      \phi(j_1) \phi(j_2) \phi(j_3) (Q_{j_2} - Q_{j_3})^{-1}
        \left( X(j_1, j_2) + X(j_1, j_3) \right) \nn
  &+& ig^2 \sum_{j_1} \sum_{j_3(\neq j_1)}
      \phi(j_1) \phi(j_1) \phi(j_3) (Q_{j_1} - Q_{j_3})^{-1}
        \left( X(j_1, j_3) \right) \nn
  &+& ig^2 \sum_{j_1} \sum_{j_2(\neq j_1)} \sum_{j_3(\neq j_2)}
      \phi(j_1) \phi(j_2) \phi(j_3) (Q_{j_1} - Q_{j_2})^{-1}
        \left( X(j_1, j_3) + X(j_2, j_3) \right) \nn
  &+& ig^2 \sum_{j_1} \sum_{j_2(\neq j_1)}
      \phi(j_1) \phi(j_2) \phi(j_2) (Q_{j_1} - Q_{j_2})^{-1}
        \left( X(j_1, j_2) \right) \nn
  &=& ig^2 \sum_{j_1} \sum_{j_2(\neq j_1)} \sum_{j_3(\neq j_2)}
      \phi(j_1) \phi(j_2) \phi(j_3) (Q_{j_1} - Q_{j_2})^{-1}
        (Q_{j_2} - Q_{j_3})^{-1} \nn
  & & \ \ \ \ \times
      \left( X(j_1, j_2) (Q_{j_1} - Q_{j_2}) + X(j_1, j_3) (Q_{j_1} - Q_{j_3})
              + X(j_2, j_3) (Q_{j_2} - Q_{j_3}) \right) \nn
  &+& ig^2 \sum_{j_1} \sum_{j_3(\neq j_1)}
      \phi(j_1) \phi(j_1) \phi(j_3) (Q_{j_1} - Q_{j_3})^{-1}
        \left( X(j_1, j_3) \right) \nn
  &+& ig^2 \sum_{j_1} \sum_{j_2(\neq j_1)}
      \phi(j_1) \phi(j_2) \phi(j_2) (Q_{j_1} - Q_{j_2})^{-1}
        \left( X(j_1, j_2) \right). \nonumber
\eeqa
Note that
\[
(k_{j_1}+k_{j_2}+k_{j_3})^2=
X(j_1, j_2)(Q_{j_1}-Q_{j_2})
+X(j_1, j_3)(Q_{j_1}-Q_{j_3})
+X(j_2, j_3)(Q_{j_2}-Q_{j_3}),
\]
\[
(2k_{j_1}+k_{j_2})^2=2X(j_1, j_2)(Q_{j_1}-Q_{j_2}).
\]
Then the minimal solution of $\Phi^{(3)}(x)$ is given by
\beqa
\Phi^{(3)}(x)&=&
-ig^2 \sum_{j_1}\sum_{j_2(\neq j_1)}\sum_{j_3(\neq j_2)}
\phi(j_1)\phi(j_2)\phi(j_3)(Q_{j_1}-Q_{j_2})^{-1}(Q_{j_2}-Q_{j_3})^{-1} \nn
& &-
ig^2 \sum_{j_1}\sum_{j_2(\neq j_1)}
\phi(j_1)\phi(j_1)\phi(j_2)\frac{1}{2}(Q_{j_1}-Q_{j_2})^{-2} \nn
& &-ig^2 \sum_{j_1}\sum_{j_2(\neq j_1)}
\phi(j_1)\phi(j_2)\phi(j_2)\frac{1}{2}(Q_{j_1}-Q_{j_2})^{-2}.
\eeqa
Similarly, we have the minimal solution for $\Phi^{(4)}(x)$:
\beqa
& &\Phi^{(4)}(x)/(-ig^3) \nn
&=&
\sum_{j_1}\sum_{j_2(\neq j_1)}\sum_{j_3(\neq j_2)}\sum_{j_4(\neq j_3)}
\phi(j_1)\phi(j_2)\phi(j_3)\phi(j_4)
(Q_{j_1}-Q_{j_2})^{-1}(Q_{j_2}-Q_{j_3})^{-1}(Q_{j_3}-Q_{j_4})^{-1}
\nn
&+&\sum_{j_1}\sum_{j_2(\neq j_1)}\sum_{j_3(\neq j_2)}
\phi(j_1)\phi(j_1)\phi(j_2)\phi(j_3)\frac{1}{2}(Q_{j_1}-Q_{j_2})^{-2}
(Q_{j_2}-Q_{j_3})^{-1} \nn
&+&\sum_{j_1}\sum_{j_2(\neq j_1)}\sum_{j_3(\neq j_2)}
\phi(j_1)\phi(j_2)\phi(j_2)\phi(j_3) \nn
& &\ \ \ \ \ \times \left(
\frac{1}{2}(Q_{j_1}-Q_{j_2})^{-2}
(Q_{j_2}-Q_{j_3})^{-1}+
\frac{1}{2}(Q_{j_1}-Q_{j_2})^{-1}
(Q_{j_2}-Q_{j_3})^{-2} 
\right)\nn
&+&\sum_{j_1}\sum_{j_2(\neq j_1)}\sum_{j_3(\neq j_2)}
\phi(j_1)\phi(j_2)\phi(j_3)\phi(j_3)\frac{1}{2}(Q_{j_1}-Q_{j_2})^{-1}
(Q_{j_2}-Q_{j_3})^{-2} \nn
&+&\sum_{j_1}\sum_{j_2(\neq j_1)}
\phi(j_1)\phi(j_1)\phi(j_1)\phi(j_2)\frac{1}{3!}(Q_{j_1}-Q_{j_2})^{-3} \nn
&+&\sum_{j_1}\sum_{j_2(\neq j_1)}
\phi(j_1)\phi(j_1)\phi(j_2)\phi(j_2)\frac{1}{2}(Q_{j_1}-Q_{j_2})^{-3} \nn
&+&\sum_{j_1}\sum_{j_2(\neq j_1)}
\phi(j_1)\phi(j_2)\phi(j_2)\phi(j_2)\frac{1}{3!}(Q_{j_1}-Q_{j_2})^{-3}.
\eeqa

Let us now consider the general case $\Phi^{(m)}(x)$. 
We define coefficients $D$ as follows:
\beqn
  \Phi^{(m)}(x) = -ig^{m - 1} \sum_{j_1 = 1}^n \ldots \sum_{j_m = 1}^n
                     \phi(j_1) \ldots \phi(j_m) D(j_1, \ldots, j_m).
\label{sAa}
\eeqn
See (\ref{phj}). 
Then the recursion relation (\ref{Phrec}) can be transformed into 
the recursion relations for $D$:
\beqa
  & & (k_{j_1} + \ldots + k_{j_m})^2 D(j_1, \ldots, j_m) \nn
  &=& \sum_{l = 1}^{m - 1} D(j_1, \ldots, j_l) D(j_{l + 1}, \ldots, j_m)
       \left( \sum_{s = 1}^l \sum_{t = l + 1}^{m}X(j_s, j_t) \right).
\label{recAa}
\eeqa
See (\ref{X12}).
It is convenient to rearrange the sum (\ref{sAa}) into the following form:
\beqa
  \Phi^{(m)} &= & -ig^{m - 1} 
     \sum_{l = 1}^{m} \sum_{j_1} \sum_{j_2(\neq j_1)} 
        \ldots \sum_{j_l(\neq j_{l - 1})}
          \sum_{n_1\geq 1} \ldots \sum_{n_l \geq 1} \nn
       & & \times \delta_{n_1 + \ldots + n_l, m}
           \phi(j_1)^{n_1} \phi(j_2)^{n_2} \ldots \phi(j_l)^{n_l}
                D(1^{n_1} 2^{n_2} \ldots l^{n_l}).
\label{PhD}
\eeqa
We introduce a notation:
\[
  D(1^{n_1} 2^{n_2} \ldots l^{n_1}) = 
    D(\overbrace{j_1, \ldots, j_1}^{n_1}, 
        \overbrace{j_2, \ldots, j_2}^{n_2}, \ldots, 
        \overbrace{j_l, \ldots, j_l}^{n_l}),
\]
\[
  X_{st} = X(j_s, j_t), \ \ \ \ Q_{st} = Q_{j_s}-Q_{j_t}.
\]
The initial condition for $m=1$ and the 
minimality of the solution are summarized by
\beqn
  D(1^{n_1}) = \delta_{n_1,1}.
\label{ini}
\eeqn
Then the recursion relations (\ref{recAa}) are given by
\beqa
  & & \sum_{s < t} n_s n_ t Q_{st} X_{st} D(1^{n_1} 2^{n_2} \ldots l^{n_l}) \nn
  &=& \sum_{s < t} \sum_{p = 1}^{n_s} p n_t X_{st} 
             D(1^{n_1} 2^{n_2} \ldots (s - 1)^{n_{s - 1}} s^p)
             D(s^{n_s - p} (s + 1)^{n_{s + 1}} \ldots l^{n_l}) \nn
  &+& \sum_{s < t} \sum_{q = s +1}^{t - 1} \sum_{p = 1}^{n_q} n_s n_t X_{st}
             D(1^{n_1} 2^{n_2} \ldots (q - 1)^{n_{q - 1}} q^p)
             D(q^{n_q - p} (q + 1)^{n_{q + 1}} \ldots l^{n_l}) \nn
  &+& \sum_{s < t} \sum_{p = 1}^{n_t - 1} n_s (n_t - p) X_{st}
           D(1^{n_1} 2^{n_2} \ldots (t - 1)^{n_{t - 1}} t^p)
             D(t^{n_t - p} (t + 1)^{n_{t + 1}} \ldots l^{n_l}).
\eeqa
The solutions $D$ do not contain the factor $X_{st}$. 
The above recursion relation is decomposed into the set of recursion
relations by comparing the coefficients at $X_{st}$ $(s < t)$.
We have recursion relations:
\beqa
  & & Q_{s, t} D(1^{n_1} 2^{n_2} \ldots l^{n_l}) \nn
  &=& \sum_{p = 1}^{n_s} \frac{p}{n_s} 
             D(1^{n_1} 2^{n_2} \ldots (s - 1)^{n_{s - 1}} s^p)
             D(s^{n_s - p} (s + 1)^{n_{s + 1}} \ldots l^{n_l}) \nn
  &+& \sum_{q = s +1}^{t - 1} \sum_{p = 1}^{n_q}
             D(1^{n_1} 2^{n_2} \ldots (q - 1)^{n_{q - 1}} q^p)
             D(q^{n_q - p} (q + 1)^{n_{q + 1}} \ldots l^{n_l}) \nn
  &+& \sum_{p = 1}^{n_t - 1} \frac{ (n_t - p) }{n_t}
           D(1^{n_1} 2^{n_2} \ldots (t - 1)^{n_{t - 1}} t^p)
             D(t^{n_t - p} (t + 1)^{n_{t + 1}} \ldots l^{n_l}).
\label{recAb}
\eeqa
Not all recursion relations (\ref{recAb}) are independent.
Note that
\[
  Q_{st} = \sum_{u = s + 1}^t Q_{u-1, u}.
\]
All $Q_{st}$ can be expressed by $Q_{u-1, u}$ $(u = 2, \ldots, l)$.
The set of recursion relations (\ref{recAb}) are generated from
\beqa
  & & Q_{t-1, t} D(1^{n_1} 2^{n_2} \ldots l^{n_l}) \nn
  &=& \sum_{p = 1}^{n_{t - 1}} \frac{p}{n_{t - 1}} 
          D(1^{n_1} 2^{n_2} \ldots (t - 2)^{n_{t - 2}} (t - 1)^p)
             D((t - 1)^{n_{t - 1} - p} t^{n_t} \ldots l^{n_l}) \nn
  &+& \sum_{p = 1}^{n_t - 1} \frac{ (n_t - p) }{n_t}
           D(1^{n_1} 2^{n_2} \ldots (t - 1)^{n_{t - 1}} t^p)
             D(t^{n_t - p} (t + 1)^{n_{t + 1}} \ldots l^{n_l}), 
\label{recXt}
\eeqa
for $t = 2, \ldots, l$.
But not all recursion relations (\ref{recXt}) are needed to determine
the coefficient $D$. With the initial condition (\ref{ini}), the
recursion relation (\ref{recXt}) for $t=2$ is sufficient to obtain the 
answer. We must check that the solution satisfies the other 
recursion relation (\ref{recXt}) for $t=3, \ldots, l$.

Combining eq.(\ref{ini}) and eq.(\ref{recXt}) for $t=2$, we have the
recursion relation:
\beqa
\label{Arec1}
  & & D(1^{n_1} 2^{n_2} \ldots l^{n_l}) \\
  &=& \frac{1}{n_1 n_2} Q_{12}^{-1}
      \left( n_2 D(1^{n_1 - 1} 2^{n_2} \ldots l^{n_l})
             + \sum_{p = 1}^{n_2-1} n_1 (n_2 - p) D(1^{n_1} 2^p)
               D(2^{n_2-p} 3^{n_3}\ldots l^{n_l}) \right) \nonumber.
\eeqa
To solve this relation, we first determine the coefficient 
$D(1^{n_1} 2^p)$.
The simplest case is 
\[
D(12)=Q_{12}^{-1}D(1)D(2)=Q_{12}^{-1}.
\]
For $n_2>1$, we have
\beqn
  D(12^{n_2}) = \frac{1}{n_2}   Q_{12}^{-1} D(12^{n_2 - 1})
              = \frac{1}{n_2!} (Q_{12}^{-1})^{n_2}.
\eeqn
Next we consider $D(1^{n_1}2^{n_2})$.
\beqn
  D(1^{n_1} 2^{n_2}) = \frac{1}{n_1n_2} Q_{12}^{-1}
      \left( n_2 D(1^{n_1 - 1} 2^{n_2}) + n_1 D(1^{n_1} 2^{n_2 - 1}) \right)
\eeqn
Let us define $C(n_1, n_2)$ by:
\beqn
  D(1^{n_1} 2^{n_2}) 
  = \frac{1}{n_1! n_2!} (Q_{12}^{-1})^{n_1 + n_2 - 1} C(n_1, n_2).
\eeqn
Then the coefficient $C(n_1, n_2)$ satisfies
\beqn
  C(n_1, n_2) = C(n_1 - 1, n_2) + C(n_1, n_2 - 1).
\eeqn
With the initial conditions $  C(1, n_2) = C(n_1, 1) = 1 $,
the solution is given by
\[
  C(n_1, n_2) = \frac{(n_1 + n_2 - 2)!}{(n_1 - 1)!(n_2 - 1)!}.
\]
Therefore $D(1^{n_1}2^{n_2})$ is determined as
\beqn
  D(1^{n_1}2^{n_2}) 
  = \frac{(n_1 + n_2 - 2)!} {n_1!(n_1 - 1)!n_2!(n_2- 1)!} 
    (Q_{12}^{-1})^{n_1 + n_2 - 1}.
\eeqn
Let us insert this relation into eq.(\ref{Arec1}). Then we obtain
\beqa
      D(1^{n_1} 2^{n_2} \ldots l^{n_l}) 
  &=& \frac{1}{n_1} Q_{12}^{-1} D(1^{n_1 - 1} 2^{n_2} \ldots l^{n_l}) \nn
  & & + \sum_{p = 1}^{n_2 - 1}
        \frac{(n_2 - p)}{n_1!\  p!\ n_2} C(n_1, p) (Q_{12}^{-1})^{n_1 + p}
         D(2^{n_2 - p} \ldots l^{n_l}).
\eeqa
Using this relation recursively, we have
\beqa
     D(1^{n_1} 2^{n_2} \ldots l^{n_l}) 
  &=& \frac{1}{n_1 !}(Q_{12}^{-1})^{n_1}D(2^{n_2} \ldots l^{n_l}) \nn
  &+& \sum_{p = 1}^{n_2 - 1} \frac{(n_2 - p)}{n_1!\  p!\  n_2} 
          \left(\sum_{s = 1}^{n_1} C(n_1 + 1 - s, p) \right)
             (Q_{12}^{-1})^{n_1 + p} D(2^{n_2 - p} \ldots l^{n_l}). 
\eeqa
With the help of the relation,
\beqn
  C(n_1, p + 1) = \sum_{s = 1}^{n_1} C(s, p),
\eeqn
we obtain the following recursion relations:
\beqa
\label{Arec2}
      D(1^{n_1} 2^{n_2} \ldots l^{n_l}) 
  &=& \sum_{p_2 = 0}^{n_2 - 1}
      \frac{( n_2 - p_2)\  C(n_1, p_2 + 1) }{ n_1!\ p_2!\  n_2 } 
         ( Q_{12}^{-1} )^{ n_1 + p_2} D(2^{n_2 - p_2} \ldots l^{n_l}) \nn
  &=& \sum_{p_2 = 1}^{n_2}
      \frac{p_2\  C(n_1, n_2 + 1 - p_2)}{ n_1!\  (n_2 - p_2)!\  n_2 } 
           ( Q_{12}^{-1} )^{ n_1 + n_2 - p_2} D(2^{p_2} \ldots l^{n_l}).
\eeqa
Here we replaced $p_2$ with $n_2-p_2$.

Then, solutions of the recursion relations
for $D$ are finally given by
\beqa
  & & D(1^{n_1} 2^{n_2} \ldots l^{n_l}) \nn
  &=& \sum_{p_2 = 1}^{n_2} \sum_{p_3 = 1}^{n_3}
                    \ldots \sum_{p_{l - 1} = 1}^{n_{l - 1}}
        \prod_{s = 1}^{l}(n_s)^{-1}\prod_{t = 2}^l
          \left(
             \frac{ (p_{t - 1} + n_t - p_t - 1)! }
                  { \left( (p_{t - 1} - 1)! (n_t - p_t)! \right)^2 }
                  ( Q_{t - 1, t}^{-1})^{p_{t - 1} + n_t - p_t}
          \right)
\label{Asol}
\eeqa
with the convention $p_1=n_1$ and $p_l=1$.
We can see that this solution (\ref{Asol}) indeed
satisfies recursion relations (\ref{recXt}). 
Combination of formulas (\ref{exg}), 
(\ref{PhD}) and (\ref{Asol}) gives the perturbative solution of self-dual
equation (\ref{SD}).

Some explicit examples of $D$ are given by:
\beqn
  D(12\ldots m) = Q_{12}^{-1} Q_{23}^{-1} \ldots Q_{m-1, m}^{-1},
\eeqn
\beqn
 D(1^{n_1}2\ldots l) =
  \frac{1}{n_1!}( Q_{12}^{-1} )^{n_1} Q_{23}^{-1} \ldots Q_{l - 1, l}^{-1}.
\eeqn

\section{Scattering Operator}

Let us simplify the expression for the solution:
\beqa
  \Phi(x)
  &=& \sum_{m = 1}^{\infty} \Phi^{(m)}(x) \nn
  &=& -i \sum_{m = 1}^{\infty} g^{m - 1} \sum_{l = 1}^m
         \sum_{j_1} \sum_{j_2 (\neq j_1)} \ldots \sum_{j_l (\neq j_{l-1})}
         \sum_{n_1 = 1}^{\infty} \ldots \sum_{n_l = 1}^{\infty} \nn
  & & \ \ \ \ \ \times \delta_{n_1 + \ldots + n_l, m}
         \phi(j_1)^{n_1} \ldots \phi(j_l)^{n_l} D(1^{n_1} \ldots l^{n_l}) \nn
  &=& 
      -i \sum_{l = 1}^{\infty} \sum_{\{j_i \neq j_{i + 1}\}} 
         \sum_{\{n_i \geq 1\}} g^{n_1 + \ldots + n_l - 1}
       \phi(j_1)^{n_1} \ldots \phi(j_l)^{n_l} D(1^{n_1} \ldots l^{n_l}).
\eeqa
See (\ref{PhD}) and (\ref{Asol}).  To obtain the last expression, we
first exchange the order of summation with respect to $m$ and
$l$. Then we use $\delta_{n_1 + \ldots + n_l, m}$ in order to perform
summation with respect to $m$.

The coefficients $D$ depend on
the momenta by means of $Q_{i-1, i}^{-1}=(Q_{j_{i-1}}-Q_{j_i})^{-1}$:
\[
  D(1^{n_1} \ldots l^{n_l})
  = D(1^{n_1} \ldots l^{n_l}; Q_{12}^{-1}, \ldots, Q_{l - 1, l}^{-1}).
\]
The factor $Q_{i - 1, i}^{-1}$ diverges for $j_{i - 1} = j_i$.
To regularize this divergence, 
we replace $Q_{i-1, i}^{-1}$ in $D$ with 
$P_{i-1, i} = P(j_{i - 1}, j_i)$. See eq.(\ref{P12}).
Then the solution of the self-dual equation can be written as
\beqn
\Phi(x)=\sum_{l=1}^{\infty}\tilde{\Phi}^{(l)}(x).
\eeqn
This is not the expansion in the coupling constant. 
$\tilde{\Phi}^{(l)}$ is not homogeneous in $g$.
Here
\beqn
  \tilde{\Phi}^{(l)}(x) = \sum_{j_1 = 1}^{n} \sum_{j_2 = 1}^{n}
    \ldots \sum_{j_l = 1}^n \tilde{\Phi}^{(l)}(j_1, \ldots, j_l),
\label{Phlx}
\eeqn
and
\beqa
  \tilde{\Phi}^{(l)}(j_1, \ldots, j_l)
  &=& -i \sum_{n_1 = 0}^{\infty} \ldots \sum_{n_l = 0}^{\infty} 
     g^{n_1 + \ldots + n_l + l - 1} \nn
  & & \ \ \ \times
       \phi(j_1)^{n_1 + 1} \ldots \phi(j_l)^{n_l + 1}
          D(1^{n_1 + 1} \ldots l^{n_l + 1}; P_{12}, \ldots, P_{l - 1, l}).
\label{PhtDj}
\eeqa
Especially for $l=1$,
\beqn
  \tilde{\Phi}^{(1)}(j_1) = -i \phi(j_1)
  = -i T^{a_{j_1}} e^{-ik_{j_1} x} f(k_j).
\eeqn

We consider $\tilde{\Phi}^{(2)}(j_1, j_2)$.
\beqn
  \tilde{\Phi}^{(2)}(j_1, j_2)
  = -ig^{-1}\sum_{n_1=0}^{\infty}\sum_{n_2=0}^{\infty}
     (g\phi(j_1))^{n_1+1}(g\phi(j_2))^{n_2+1}
     \frac{(n_1+n_2)!}{(n_1+1)!n_1!(n_2+1)!n_2!} (P_{12})^{n_1+n_2+1}.
\label{Pht2}
\eeqn
Note that $\phi(j_1)$ takes a value in the Lie algebra of the gauge
group. The order of $\phi$'s needs to be considered.

We shall use the integral representation for the Gamma function
\beqn
  m!=\int_0^{\infty}ds \ e^{-s}\ \ s^m,
\label{intG}
\eeqn
to represent $(n_1 + n_2)!$ in (\ref{Pht2}).
We shall obtain
\beqn
\tilde{\Phi}^{(2)}(j_1, j_2)
=-i g^{-1} \sum_{n_1=0}^{\infty}\sum_{n_2=0}^{\infty}
\int_0^{\infty}ds\ e^{-s}\frac{
s^{n_1+n_2}(g\phi(j_1))^{n_1+1}(g\phi(j_2))^{n_2+1}
(P_{12})^{n_1+n_2+1}}
{(n_1+1)!n_1!(n_2+1)!n_2!}.
\eeqn
Now the summation in $n_1$ and $n_2$ factorizes.

The sum in $n_1$ or $n_2$ in the above equation can be expressed
in terms of:
\beqn
     V_1(a) = \sum_{n=0}^{\infty}\frac{1}{(n+1)!n!}a^n 
            = \int_0^1d\alpha\ V(\alpha a).
\label{V1}
\eeqn
Here $V(a)$ is given by eq.(\ref{mB0}).
Note that $V_1(a)$ has several representations:
\beqn
  V_1(a) = \oint \frac{dt}{2\pi i}e^{1/t+at}
         = I_1(2\sqrt{a})/\sqrt{a}
         = \frac{d}{da}V(a).
\eeqn
Here $I_1$ is the modified Bessel function of the first kind.
The integration contour is a circle in the complex plane around zero in
positive direction.

Now the solution $\tilde{\Phi}^{(2)}(j_1, j_2)$ can be expressed as
\beqn
\tilde{\Phi}^{(2)}(j_1, j_2)=-i
\int_0^1 d\alpha_1 \int_0^1 d\alpha_2
T(\alpha_1, \alpha_2; j_1, j_2)\phi(j_2),
\eeqn
where $T(\alpha_1, \alpha_2; j_1, j_2)$ is defined by
\beqa
  & &T(\alpha_1, \alpha_2; j_1, j_2) \nn
  &=&
    g \phi(j_1) P(j_1, j_2)
       \int_0^{\infty}ds\ e^{-s}
         V\Bigl( s \alpha_1 g \phi(j_1) P(j_1, j_2) \Bigr)
         V\Bigl( s \alpha_2 g \phi(j_2) P(j_1, j_2) \Bigr).
\label{scop2}
\eeqa
Then $\tilde{\Phi}^{(2)}(x)$ is given by
\beqn
\label{Pht}
\tilde{\Phi}^{(2)}(x)=-i\sum_{j_1=1}^{n}\sum_{j_2=1}^{n}
\int_0^1 d\alpha_1 \int_0^1 d\alpha_2
T(\alpha_1, \alpha_2; j_1, j_2)\phi(j_2).
\eeqn
The integration variable $\alpha$ was introduced by formula (\ref{V1}).

Using a similar technique, we can write the coefficient $D$ 
(\ref{Asol}) as follows:
\beqa
& &D(1^{n_1+1}2^{n_2+1}\ldots l^{n_l+1}) \nonumber \\
&=&\sum_{p_2=0}^{n_2}\ldots\sum_{p_{l-1}=0}^{n_{l-1}}
\prod_{t=2}^{l}
\int_0^{\infty}ds_t\ e^{-s_t}P_{t-1, t} \nonumber \\
& &\times
\frac{(s_2P_{12})^{n_1}}{(n_1+1)!n_1!}
\prod_{t=2}^{l-1}
\frac{(s_tP_{t-1, t})^{n_t-p_t}(s_{t+1}P_{t, t+1})^{p_t}}
{(n_t+1)(p_t!(n_t-p_t)!)^2}
\frac{(s_lP_{l-1, l})^{n_l}}{(n_l+1)!n_l!}.
\eeqa
We replaced $p_t$ with $p_t + 1$ and 
represented $(p_{t - 1} + n_t - p_t)!$ by $s_t$-integration.
In order to calculate $\tilde{\Phi}^{(l)}$,
we need to substitute this expression for $D$ into the sum (\ref{PhtDj}).
Then we have
\beqa
  & & \tilde{\Phi}^{(l)}(j_1, \ldots, j_l) \nn
  &=& -i \sum_{n_1 = 0}^{\infty} \ldots \sum_{n_l = 0}^{\infty}
         \sum_{p_2 = 0}^{n_2} \ldots \sum_{p_{l - 1} = 0}^{n_{l - 1}}
      \prod_{t = 2}^l \int_0^{\infty} ds_t\ e^{-s_t}
      g\phi(j_1)P_{12}
      \frac{(s_2 g \phi(j_1) P_{12})^{n_1}}{(n_1+1)!n_1!} \nn
  & & \times
       \prod_{t=2}^{l-1} 
         \frac{(s_t g \phi(j_t) P_{t - 1, t})^{n_t-p_t}
              g \phi(j_t) P_{t, t+1} (s_{t+1} g \phi(j_t) P_{t, t+1})^{p_t}}
                {(n_t+1)(p_t!(n_t-p_t)!)^2}
         \frac{(s_l g \phi(j_l) P_{l-1, l})^{n_l}}{(n_l+1)!n_l!} \phi(j_l).
\eeqa
Using formula (\ref{V1}), we introduce integration over $\alpha_1$ and
$\alpha_l$ and perform the summation in $n_1$ and $n_l$:
\beqa
  & & \tilde{\Phi}^{(l)}(j_1, \ldots, j_l) \nn
  &=& -i \sum_{n_2 = 0}^{\infty} \ldots \sum_{n_{l - 1} = 0}^{\infty}
         \sum_{p_2 = 0}^{n_2} \ldots \sum_{p_{l - 1} = 0}^{n_{l - 1}}
      \prod_{t = 2}^l \int_0^{\infty} ds_t\ e^{-s_t}
      \int_0^1 d\alpha_1 \int_0^1 d\alpha_l
      g\phi(j_1)P_{12}
      V(\alpha_1 s_2 g \phi(j_1) P_{12}) \nn
  & & \times
       \prod_{t=2}^{l-1} 
         \frac{(s_t g \phi(j_t) P_{t-1, t})^{n_t-p_t}
              g \phi(j_t) P_{t, t + 1} (s_{t+1} g \phi(j_t) P_{t, t+1})^{p_t}}
                {(n_t+1)(p_t!(n_t-p_t)!)^2}
         V(\alpha_l s_l g \phi(j_l) P_{l-1, l}) \phi(j_l).
\eeqa
In order to calculate summation in $n_t$ and $p_t$ 
for $j = 2, \ldots, l - 1$, we need to calculate the sum:
\beqn
\label{sumd}
  \sum_{n_t = 0}^{\infty} \sum_{p_t = 0}^{n_t}
    \frac{(s_t g \phi(j_t)P_{t-1, t})^{n_t-p_t}
       g \phi(j_t)P_{t, t + 1}
      (s_{t+1} g \phi(j_t) P_{t, t+1})^{p_t}} {(n_t+1)(p_t!(n_t-p_t)!)^2}.
\eeqn
We first exchange the order of summation then replace $n_t$ with $n_t + p_t$:
\beqa
  & &\sum_{p_t = 0}^{\infty} \sum_{n_t = p_t}^{\infty}
    \frac{(s_t g \phi(j_t)P_{t-1, t})^{n_t-p_t}
       g \phi(j_t)P_{t, t + 1}
      (s_{t+1} g \phi(j_t) P_{t, t+1})^{p_t}}
        {(n_t+1)(p_t!(n_t-p_t)!)^2} \nn
  &=&\sum_{p_t = 0}^{\infty} \sum_{n_t = 0}^{\infty}
    \frac{(s_t g \phi(j_t)P_{t-1, t})^{n_t}
       g \phi(j_t)P_{t, t + 1}
      (s_{t+1} g \phi(j_t) P_{t, t+1})^{p_t}}
        {(n_t + p_t + 1)(p_t! n_t!)^2}.
\eeqa
Let us represent $(n_t + p_t + 1)^{-1}$ by
\beqn
  \frac{1}{n_t + p_t + 1}=\int_0^1 d\alpha_t\ \alpha_t^{n_t + p_t}.
\eeqn
This formula introduces $\alpha_t$-integration. The sum 
in (\ref{sumd}) can be written as
\beqn
  \int_0^1 d\alpha_t 
    \left( \sum_{n_t = 0}^{\infty} 
       \frac{(\alpha_t s_t g \phi(j_t) P_{t - 1, t})^{n_t}}
            {(n_t!)^2} \right) g \phi(j_t) P_{t, t + 1}
    \left( \sum_{p_t = 0}^{\infty} 
       \frac{(\alpha_t s_{t+1} g \phi(j_t) P_{t, t + 1} )^{p_t}}
             {(p_t!)^2} \right).
\eeqn
By means of (\ref{mB0}) the sum (\ref{sumd}) is equivalent to
\[
  \int_0^1 d\alpha_t
    V(\alpha_t s_t g \phi(j_t) P_{t-1, t}) 
    g \phi(j_t) P_{t, t + 1} V(\alpha_t s_{t+1} g \phi(j_t) P_{t, t + 1}).
\]
Then, we have
\beqa
  & &\tilde{\Phi}^{(l)}(j_1, j_2, \ldots, j_l) \nn
  &=& - i\int_0^1 d\alpha_1 \int_0^1 d\alpha_2 \ldots
           \int_0^1 d\alpha_l \prod_{t=2}^l  \int_0^{\infty} d s_t
              e^{-s_t} \
     g\phi(j_1)P_{12} V(\alpha_1 s_2 g \phi(j_1) P_{12})  \nn
  & & \ \ \ \ \times
    \prod_{t = 2}^{l - 1}
    \Bigl(V(\alpha_t s_t g \phi(j_t) P_{t-1, t}) 
    g \phi(j_t) P_{t, t + 1} 
     V(\alpha_t s_{t+1} g \phi(j_t) P_{t, t + 1}) \Bigr) 
    V(\alpha_l s_l g \phi(j_l) P_{l-1, l}) \phi(j_l) \nn
  &=& - i\int_0^1 d\alpha_1 \int_0^1 d\alpha_2 \ldots
              \int_0^1 d\alpha_l \nn
  & & \ \ \ \ \times
    \prod_{t = 2}^l
    \Bigl( \int_0^{\infty} d s_t e^{-s_t} g \phi(j_{t - 1}) P_{t - 1 , t}
    V(\alpha_{t - 1} s_t g \phi(j_{t - 1}) P_{t-1, t}) 
    V(\alpha_t s_t g \phi(j_t) P_{t-1, t}) \Bigr)\phi(j_l). \nonumber
\eeqa
Recalling the definition of $T$ (\ref{scop2}), we have
\beqa
  \tilde{\Phi}^{(l)}(j_1, j_2, \ldots, j_l)
  &=& - i\int_0^1 d\alpha_1 \int_0^1 d\alpha_2 \ldots
              \int_0^1 d\alpha_l \nn
  & & \ \ \ \ \times
   T(\alpha_1, \alpha_2; j_1, j_2) T(\alpha_2, \alpha_3; j_2, j_3)
      \ldots T(\alpha_{l-1}, \alpha_l; j_{l-1}, j_l)\phi(j_l).
\eeqa
Therefore, we have
\beqa
  \tilde{\Phi}^{(l)}(x)
  &=& -i \sum_{j_1 = 1}^n \sum_{j_2 = 1}^n
                   \ldots \sum_{j_l = 1}^n
         \int_0^1 d\alpha_1 \int_0^1 d\alpha_2 
                     \ldots \int_0^1 d\alpha_l \nn
  & &    \ \ \ \ \times
         T(\alpha_1, \alpha_2; j_1, j_2)T(\alpha_2, \alpha_3; j_2, j_3)
             \ldots T(\alpha_{l - 1}, \alpha_l; j_{l - 1}, j_l) \phi(j_l).
\eeqa
See (\ref{Phlx}).
Using the `scattering operator',
 we obtain a simple representation for $\tilde{\Phi}^{(l)}$,
\beqn
  \tilde{\Phi}^{(l)}(x) = - i \vec{\phi}_0 \cdot (\hat{T})^{l-1} \vec{\phi}.
\eeqn
See (\ref{tvec}). Thus, we proved our main formula (\ref{main}):
\beqn
  \Phi(x) = -i \vec{\phi}_0 \cdot 
             \left( 1 + \sum_{l = 1}^{\infty}(\hat{T})^l \right) \vec{\phi}
          = -i \vec{\phi}_0 \cdot 
             \left( \frac{1}{1 - \hat{T}} \right) \vec{\phi}.
\eeqn

\section{Summary}

Introduction of an auxiliary linear space is typical for the solution of 
two-dimensional completely integrable differential equations 
\cite{AS, FT}.
In this sense, our solution is represented similar to solutions of 
two-dimensional classical completely integrable differential equations.

\section*{Acknowledgments}
We wish to thank Professor T. Inami for useful discussions. This work
is partly supported by the National Science Foundation (NSF) under
Grants No. PHY-9321165 and the Japan Society for the Promotion of Science.
T. O. is supported by the JSPS Research Fellowships for Young Scientists.


\newpage
\setlength{\baselineskip}{13pt}

\end{document}